\documentclass[aps,prd,preprint,superscriptaddress,tightenlines,nofootinbib]{revtex4}

\usepackage{graphicx}
\usepackage{dcolumn}
\usepackage{bm}

\begin{document}

\preprint{CLNS 08/2024}       
\preprint{CLEO 08-07}         

\title{\boldmath Inclusive $\chi_{bJ}(nP)$ Decays to $D^0 X$}


\author{R.~A.~Briere}
\author{T.~Ferguson}
\author{G.~Tatishvili}
\author{H.~Vogel}
\author{M.~E.~Watkins}
\affiliation{Carnegie Mellon University, Pittsburgh, Pennsylvania 15213, USA}
\author{J.~L.~Rosner}
\affiliation{Enrico Fermi Institute, University of
Chicago, Chicago, Illinois 60637, USA}
\author{J.~P.~Alexander}
\author{D.~G.~Cassel}
\author{J.~E.~Duboscq}
\author{R.~Ehrlich}
\author{L.~Fields}
\author{R.~S.~Galik}
\author{L.~Gibbons}
\author{R.~Gray}
\author{S.~W.~Gray}
\author{D.~L.~Hartill}
\author{B.~K.~Heltsley}
\author{D.~Hertz}
\author{J.~Kandaswamy}
\author{D.~L.~Kreinick}
\author{V.~E.~Kuznetsov}
\author{H.~Mahlke-Kr\"uger}
\author{D.~Mohapatra}
\author{P.~U.~E.~Onyisi}
\author{J.~R.~Patterson}
\author{D.~Peterson}
\author{D.~Riley}
\author{A.~Ryd}
\author{A.~J.~Sadoff}
\author{X.~Shi}
\author{S.~Stroiney}
\author{W.~M.~Sun}
\author{T.~Wilksen}
\affiliation{Cornell University, Ithaca, New York 14853, USA}
% ...
\collaboration{CLEO Collaboration}
\noaffiliation


\date{July 22, 2008}

\begin{abstract} 
Using $\Upsilon(2S)$ and $\Upsilon(3S)$ data collected with the CLEO III 
detector we have searched for decays of $\chi_{bJ}$ to final states 
with open charm.  
We fully reconstruct $D^0$ mesons with $p_{D^0} > 2.5$ GeV/$c$ 
in three decay modes 
($K^-\pi^+$, $K^-\pi^+\pi^0$, and $K^-\pi^-\pi^+\pi^+$) 
in coincidence with radiative transition photons that tag 
the production of one of the $\chi_{bJ}(nP)$ states.  
Significant signals are obtained for the two $J=1$ states.  
Recent non-relativistic QCD (NRQCD) calculations 
of $\chi_{bJ}(nP) \to c\bar{c} X$ 
depend on one non-perturbative parameter per $\chi_{bJ}$ triplet.  
The extrapolation from the observed $D^0 X$ rate over a limited momentum 
range to a full $c\bar{c} X$ rate also depends on these same parameters.  
Using our data to fit for these parameters, we extract results 
which agree well with NRQCD predictions, confirming the expectation 
that charm production is largest for the $J=1$ states.  
In particular, for $J=1$, our results are consistent with $c\bar{c}g$ 
accounting for about one-quarter of all hadronic decays.  
\end{abstract}

\maketitle


\section{Introduction}

The six known $\chi_{bJ}(nP)$ $P$-wave bound states 
of a bottom quark ($b$) and its antiparticle $\bar{b}$ 
are labeled by their total angular momentum $J=0 ,1, 2$ 
and radial quantum number $n=1, 2$.  
Their decays provide a place to test predictions based on 
Quantum Chromodynamics (QCD), which describes the strong interaction 
between quarks in the Standard Model of particle physics.  
While strong coupling prevents QCD at low energies from 
being treated with naive perturbation theory, specialized 
calculational techniques have been developed and applied 
with general success.  
In the $b\bar{b}$ system of states, one can study both transitions 
among the various quantum states, which also include the $S$-wave 
$\Upsilon$ states, or else study decays which are initiated 
by annihilation of the quark-antiquark pair.  
Although the $\chi_{bJ}$ states have been known for many years 
and there have been several studies of their transitions to other 
bound states in the $b\bar{b}$ system, 
there are no published annihilation decay branching fractions.  
This Article reports the first observation of some of the 
inclusive decays of the $\chi_{bJ}(1P,2P)$ to $D^0$ mesons.  

In practice, one studies $\chi_{bJ}$ produced via the radiative transitions 
$\Upsilon(mS) \to \gamma \chi_{bJ}(nP)$ 
from $\Upsilon$ mesons produced directly at $e^+e^-$ colliders.  
The transition photons are typically used to tag $\chi_{bJ}$ events.  
Most of the $\chi_{bJ}$ radiative decays to the 
$\Upsilon$ states are well-measured \cite{pdg}; 
the largest branching fraction is quite substantial, about 35\%.   
Small, ${\cal O}(1\%)$, hadronic transitions to other 
bottomonium states, $\chi_{b1,2}(2P) \to \pi\,\pi\, \chi_{b1,2}(1P)$ 
and $\chi_{b1}(2P) \to \omega\, \Upsilon(1S)$, 
have recently been observed \cite{CLEOchihad}.  
The remainder of the decays are expected to be dominated 
by $b\bar{b}$ annihilation.  
Positive $C$-parity forbids decays via a single photon; 
the leading process is annihilation into two gluons.  
For the $J=1$ state, decay into two on-shell gluons 
is forbidden \cite{Barb76}; instead, this state decays 
preferentially via $q\bar{q}g$.  
While the $J=0,2$ decay widths are dominated by this $gg$ 
process, they also have a small admixture of $q\bar{q}g$.  

We observe $b\bar{b}$ annihilation as a decay into lighter hadrons 
and are seeking to determine whether production of charm hadrons 
is suppressed or not.  
It is well-known that in continuum hadronization 
($e^+ e^- \to \gamma \to q\bar{q}$) that charm is not suppressed, 
while in $ggg$ decays of the $\Upsilon(1S)$, an upper limit on $D^{*+}$ 
production of ${\cal B}(\Upsilon(1S) \to ggg \to D^{*+} X) < 1.9\%$ (90\% CL) 
indicates significant suppression \cite{ARGUS}.   

The earliest calculations of inclusive charm ($c\bar{c}X$) 
production from bottomonia focused on $\Upsilon \to ggg$ decays, 
giving estimates of a few percent \cite{Fritzsch78}.  
It was soon pointed out that while production of $c\bar{c}X$ 
is predicted to be suppressed in $gg$ hadronization, it is not 
expected to be suppressed in $q\bar{q}g$ hadronization \cite{Barb79}.  
Since the $gg$ process is absent for the $\chi_{b1}(nP)$ states, 
they should have higher branching fractions to $c\bar{c}X$.   
These first calculations exhibited infrared divergences manifested as 
logarithms of the binding energy which were estimated in terms of 
a confinement radius.  
The predicted ratios of branching fractions are \cite{Barb79} 
$R^{(c)}_J \equiv  {\cal B}(\chi_{bJ} \to gg, q\bar{q}g \to c\bar{c}X)
         / {\cal B}(\chi_{bJ} \to gg, q\bar{q}g) = 6\%, 25\%$, and $12\%$ 
for the $J=0, 1$, and $2$ states, respectively.  
The predictions were independent of the radial quantum number, $n$.  
The 25\% branching fraction for $J=1$ corresponds to equal rates 
for all accessible quark flavors $q$ in $q\bar{q}g$.  

With the development of non-relativistic QCD (NRQCD) techniques 
\cite{NRQCD92}, a proper treatment of the infrared divergences was 
given and thus much improved calculations became possible.  
However, initial work \cite{NRQCD95} on bottomonium decays 
approximated final-state quarks as massless.  
Recently, this was remedied, and detailed NRQCD calculations 
of massive charm production in $\chi_{bJ}$ decay have been performed 
\cite{Bodwin07}.  
Decay rates are expressed in terms of one non-perturbative parameter 
per $\chi_{bJ}$ triplet: 
$\rho_8 \equiv m_b^2 \langle{\cal O}_8\rangle / \langle{\cal O}_1\rangle$ 
where ${\cal O}_1$ (${\cal O}_8$) is a particular color-singlet 
(color-octet) four-quark operator \cite{NRQCD95,Bodwin07} 
and $m_b$ is the one-loop pole mass, $m_b \simeq 4.6$ GeV/c$^2$.  
All of the $n$-dependence in these calculations is contained in $\rho_8$, 
and $R^{(c)}_J$ is found to increase monotonically with increasing $\rho_8$.  
For illustrative purposes, we choose a common nominal value of 
$\rho_8 = 0.10$, which gives $R^{(c)}_J = 5\%, 23\%$, and $8\%$ 
for the $J=0, 1$,and $2$ states, respectively.  
These results are in general agreement with the older calculation cited 
above.  In particular, charm production is expected to be largest 
for the $J=1$ states.  
Not only the predicted $R^{(c)}_J$, but also the efficiency of our 
applied $D^0$ momentum cut, depend on $\rho_8$.  
We thus fit for  $\rho_8$ in the context of the NRQCD results in order 
to interpret the consistency of our results with theory.

To summarize, we observe charm production by observing $D^0$ 
mesons in $\chi_{bJ}$ decays.  We thereby hope to test predictions 
for the branching fractions, especially the expectation that 
the largest branching fractions will come from the $J=1$ states 
due to the dominance of $q\bar{q}g$ decays when $gg$ is absent.  
Sections II-VII present our experimental results for inclusive 
decays of $\chi_{bJ}$ to $D^0 X$, with a $D^0$ momentum cut.  
Section VIII makes the connection between these measurements 
and the theoretically-predicted total rate of $c\bar{c}X$ 
production, $R^{(c)}_J$.  Section IX summarizes our conclusions.

\section{The CLEO III Experiment and Data Sets}

We use data collected with the CLEO III detector 
\cite{cleo} at the Cornell Electron Storage Ring (CESR).
Charged particle tracking is provided by a four-layer silicon tracker 
and a 47-layer drift chamber \cite{chamb} covering 93\% of the solid angle.  
Particle identification (PID) is performed via specific ionization 
measurements ($dE/dx$) in the drift chamber supplemented by a Ring-Imaging 
Cherenkov detector (RICH) \cite{pid} which covers 80\% of the solid angle.  
Photons are detected using an electromagnetic calorimeter consisting 
of 7784 CsI(Tl) crystals \cite{cryst}. 
All of these detector elements are immersed in a 1.5 T solenoidal 
magnetic field. 

We use CLEO III data samples of 0.65, 1.27, and 1.40 fb$^{-1}$ 
at the $\Upsilon(1S), \Upsilon(2S)$, and $\Upsilon(3S)$ resonances, 
corresponding to 13.0, 9.4, and 6.1 million $\Upsilon$ mesons produced, 
respectively.  
In addition, data were also collected about 25 MeV below each resonance: 
we analyze 0.14, 0.43, and 0.16~fb$^{-1}$ from below the 
$\Upsilon(1S), \Upsilon(2S)$, and $\Upsilon(3S)$ resonances, respectively.  
We do not use a direct off-resonance subtraction, but rather use 
these samples to constrain background shapes.

\section{Experimental Technique}

This analysis includes all six known $\chi_{bJ}(nP)$ states: 
$J = 0, 1$, and $2$ and $n = 1$ and $2$.  
The $\chi_{bJ}$ states produced in radiative $\Upsilon$ decays are tagged 
by transition photons from $\Upsilon \to \gamma \chi_{bJ}$ decays; 
the $\chi_{bJ}$ yields are obtained from fits to $E_\gamma$ spectra.  
We then fit $E_\gamma$ spectra from events with a $D^0$ candidate 
in the signal mass region, using $D^0$ mass sidebands to 
remove combinatorial background under the $D^0$ signal peak.  
After correcting for $D^0$ efficiencies and branching fractions, 
the ratio of these two inclusive yields determines the fraction of $\chi_{bJ}$ 
decays with a true $D^0$ (above our $D^0$ minimum momentum requirement).  
The photon efficiencies, numbers of initial $\Upsilon(nS)$, 
and many associated systematic uncertainties largely cancel.

We finally apply some small corrections to obtain the rate 
for {\it direct} production of $D^0$ mesons in $\chi_{bJ}$ decays.  
Direct denotes the exclusion of charm production in decays 
of other bottomonium states produced by tranistions from our 
initial $\chi_{bJ}$ (for example, via $\gamma, \pi\pi, \omega$ 
transitions).  
Our focus is on direct $D^0$ production via hadronization 
of $\chi_{bJ} \to gg, q\bar{q}g$ decays only, and not on transitions 
to other $b\bar{b}$ states which subsequently decay to $D^0 X$.

\section{Event Selection}

We first select events with transition photon candidates 
with energies between $3.50 <$ ln($E_\gamma$ [MeV]) $< 5.70$ 
($33 <  E_\gamma < 299$ MeV).  
Only showers in the barrel calorimeter, $|\cos\theta|< 0.8$, 
that are isolated from charged tracks are considered.  
Hadronic shower fragments are suppressed by 
vetoing any candidate photon shower that has a charged track
pointing anywhere in the candidate's ``connected region'': 
this is a contiguous group of adjacent crystals with 
the energy deposition in each crystal, $E_{xtal}$, satisfying 
$E_{xtal} > 10$ MeV.  
An additional requirement on the fraction of energy deposited in the central 
$3\times 3$ square of a $5\times 5$ square, $E9/E25$, is applied.  
We use an energy-dependent $E9/E25$ criterion to select soft transition 
photon candidates, while photons later used in forming $\pi^0$ 
candidates, both as a veto and as $D^0$ decay daughters, 
must satisfy the requirement of $E9/E25 > 0.85$.  

Photon background in the $\Upsilon \to \gamma \chi_{bJ}$ transitions 
is dominated by $\pi^0$ decay products.  
To suppress this background, we reject photon candidates 
that, when combined with any other photon, form a $\pi^0$ candidate 
that has an invariant mass within three standard 
deviations of the nominal $\pi^0$ mass and a lab-frame opening 
angle between the two photons satisfying 
$|\cos\theta_{\gamma \gamma}| > 0.7$.  

For $D^0$ reconstruction, we select well-measured tracks consistent with 
originating from the interaction point.   
These tracks must have an impact parameter of less than 5 cm 
with respect to the interaction point along the beam direction, 
and less than 5 mm with respect to it in the transverse plane.  
Charge-conjugate final states, $\bar{D}^0 X $, are also included 
and are implied in the remainder of the paper.  
Candidate $D^0$ mesons are reconstructed via three decay modes: 
$K^-\pi^+, K^-\pi^+\pi^0$, and $K^-\pi^-\pi^+\pi^+$.  
For charged pion and kaon selection, particle identification 
combines RICH measurements with $dE/dx$ in a momentum-dependent manner.  
The $dE/dx$ information is expressed as $\sigma^{dE}_{\pi,K}$, 
the number of standard deviations between measured and expected 
ionization for the $\pi,K$ hypothesis.  
The track-dependent $dE/dx$ resolution used to normalize 
$\sigma_{\pi,K}$ includes dependencies on velocity,
$\cos\theta$, and the number of hits used for $dE/dx$.  
RICH information is characterized with a likelihood $L$; 
we use ${\cal L}_{\pi,K}$ as shorthand for $-2 \ln L_{\pi,K}$.  
When used, the RICH information is combined with $dE/dx$ into 
one combined separation variable as:  
$\Delta \chi^2_{\pi,K} = {\cal L}_{\pi,K} - {\cal L}_{K,\pi} 
+ (\sigma^{dE}_{\pi,K})^2 - (\sigma^{dE}_{K,\pi})^2$.  
The first (second) subscript is chosen for $\pi$ ($K$) identification.  
We also impose requirements on the number of detected Cherenkov photons, 
$n^{\pi,K}_\gamma$, for either the $\pi$ or $K$ hypothesis in the RICH detector.

Momentum dependence in the use of the RICH is motivated by the Cherenkov 
threshold for kaons and the need for tracks to have sufficient 
transverse momentum to reach the RICH detector given their curvature 
in the magnetic field.  
All pion candidates must satisfy $|\sigma^{dE}_\pi| < 3$.  
Pion candidates with $p < 0.50$ GeV/$c$ are accepted with that 
criteria alone, but additional requirements are added 
for some higher-momentum candidates. 
If $0.50 < p < 0.65$~GeV/$c$ and $n_\gamma^\pi > 2$, 
we also require $\Delta \chi^2_\pi < 0$.  
Candidates with $p > 0.65$ GeV/$c$ must satisfy both 
$n_\gamma^\pi > 2$ and $\Delta \chi^2_\pi < 0$.

Kaons are identified in an analogous manner to pions, 
with three additional criteria.  
First, kaon candidates must satisfy $p > 0.18$ GeV/$c$.  
Kaons lose more energy in the inner detector than pions, and 
tightly curling tracks are poorly reconstructed.  
Second, if the track momentum is greater than 0.60 GeV/$c$, then the track 
must also be within the RICH fiducial region, $|\cos\theta| < 0.80$; 
this ensures good rejection of the more numerous pions 
as the $dE/dx$ separation degrades.  
Finally, when RICH information is available, 
a tighter criterion, $\Delta \chi^2_K < -10$, 
is used compared to that employed for pions due to the relative 
abundance of pions over kaons.  

The $\pi^0$ meson candidates from $D^0 \to K^- \pi^+ \pi^0$  are 
reconstructed from pairs of photons with an invariant mass within 2.5 standard 
deviations of the nominal $\pi^0$ mass.  These candidates are 
then kinematically constrained to the $\pi^0$ mass. 
For the $K^-\pi^+\pi^0$ mode, the precision is improved with 
an additional requirement on the candidate's location 
in the Dalitz plot.  Our criteria retains the 70\% of decays 
from the most densely-populated regions of phase space 
(based on previous measurements \cite{CLEOBerg}).  

\begin{figure}
\begin{center}
\includegraphics[width=3.0in,height=3.0in]{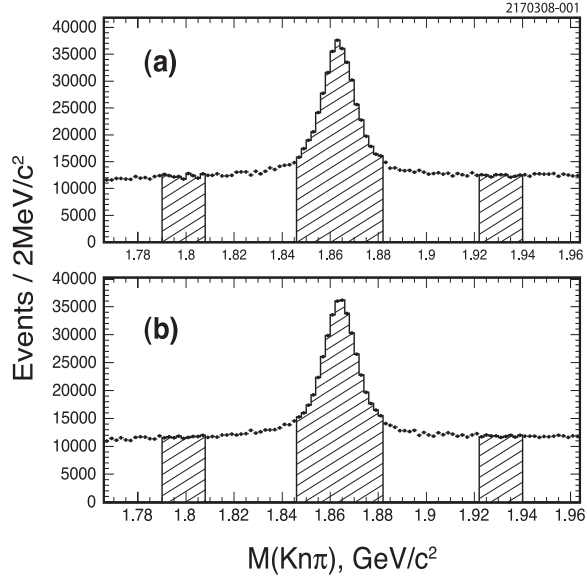}
\caption{\label{fig:md0} Sum of $K^-\pi^+$, $K^-\pi^+\pi^0$, 
  and $K^-\pi^-\pi^+\pi^+$ invariant mass distributions obtained 
  for $\Upsilon(2S)$ (a) and $\Upsilon(3S)$ (b) data.  
  The shaded areas correspond to the signal region and the two background 
  side-band regions defined in the text.}
\end{center}
\end{figure}

In order to avoid the large combinatorial backgrounds under the $D^0$ signal 
at lower momenta, only candidate $D^0$ momenta $p_{D^0} > 2.5$ GeV/$c$ 
are accepted.  
Figure~\ref{fig:md0} shows the sum of the $K^-\pi^+$, $K^-\pi^+\pi^0$, 
and $K^-\pi^-\pi^+\pi^+$ invariant mass distributions, 
$Kn\pi$ ($n=1,2,3$), obtained from $\Upsilon(2S)$ and $\Upsilon(3S)$ data 
for events also containing transition photon candidates.  
The $D^0$ signal region is defined as the 
$K^-\pi^+$, $K^-\pi^+\pi^0$, and $K^-\pi^-\pi^+\pi^+$ 
invariant mass interval $\pm 2.5 \sigma_m$ 
(using a mode-averaged $\sigma_m \simeq$ 0.0075 GeV/$c^2$) 
from the nominal $D^0$ mass, $m_{D^0}$ \cite{pdg}. 
The $D^0$ ``sideband'' regions, each with a width of $2.5 \sigma_m$, 
are located symmetrically, between 7.5 $\sigma_m$ and 10.0 $\sigma_m$ 
on either side of the nominal $D^0$ mass.

\section{Fits to the photon energy spectra}

\begin{figure}
\parbox{0.48\textwidth}
{
 \includegraphics*[width=2.7in,height=2.5in]{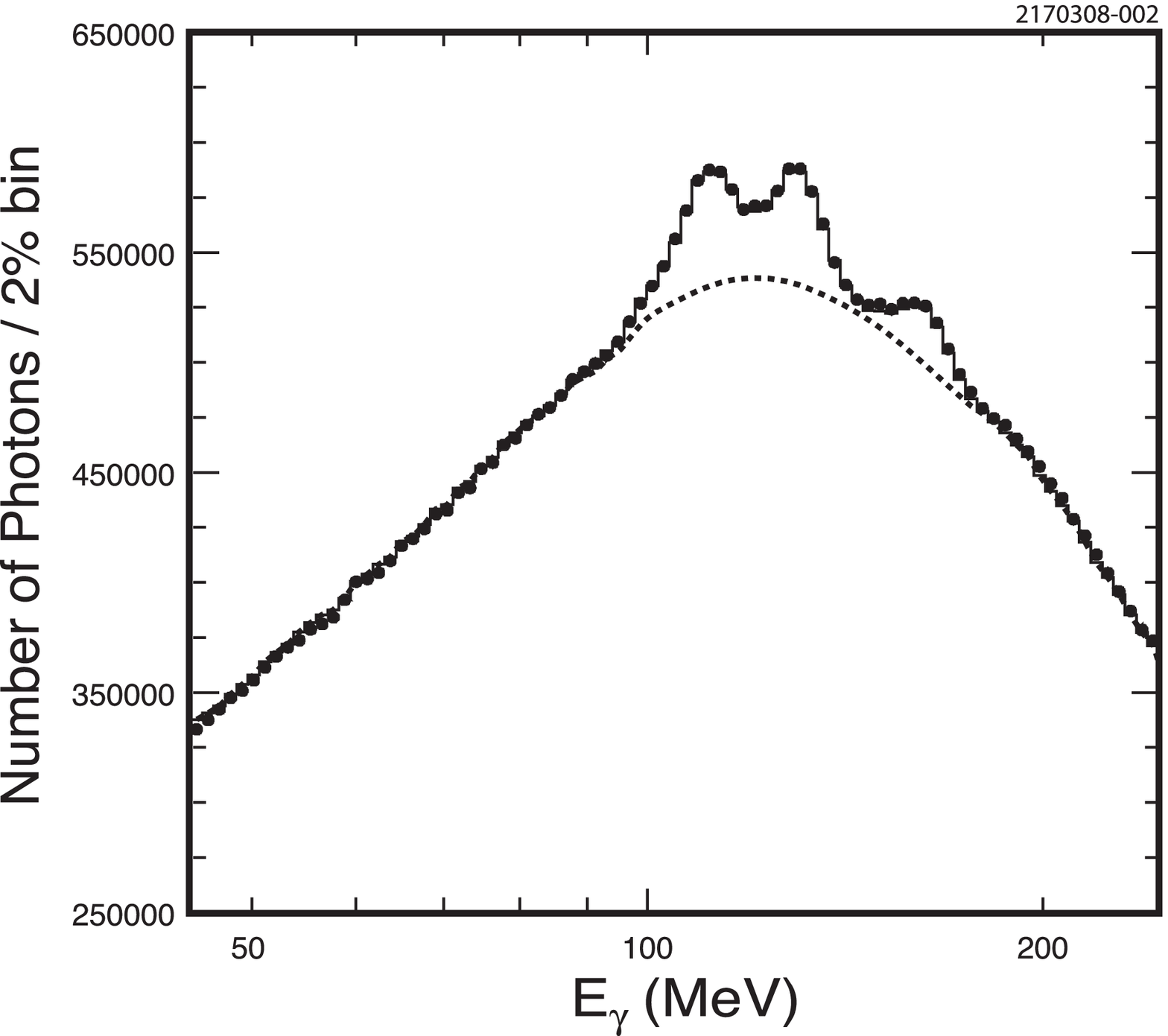}
}
\parbox{0.48\textwidth}
{
 \includegraphics*[width=2.7in,height=2.5in]{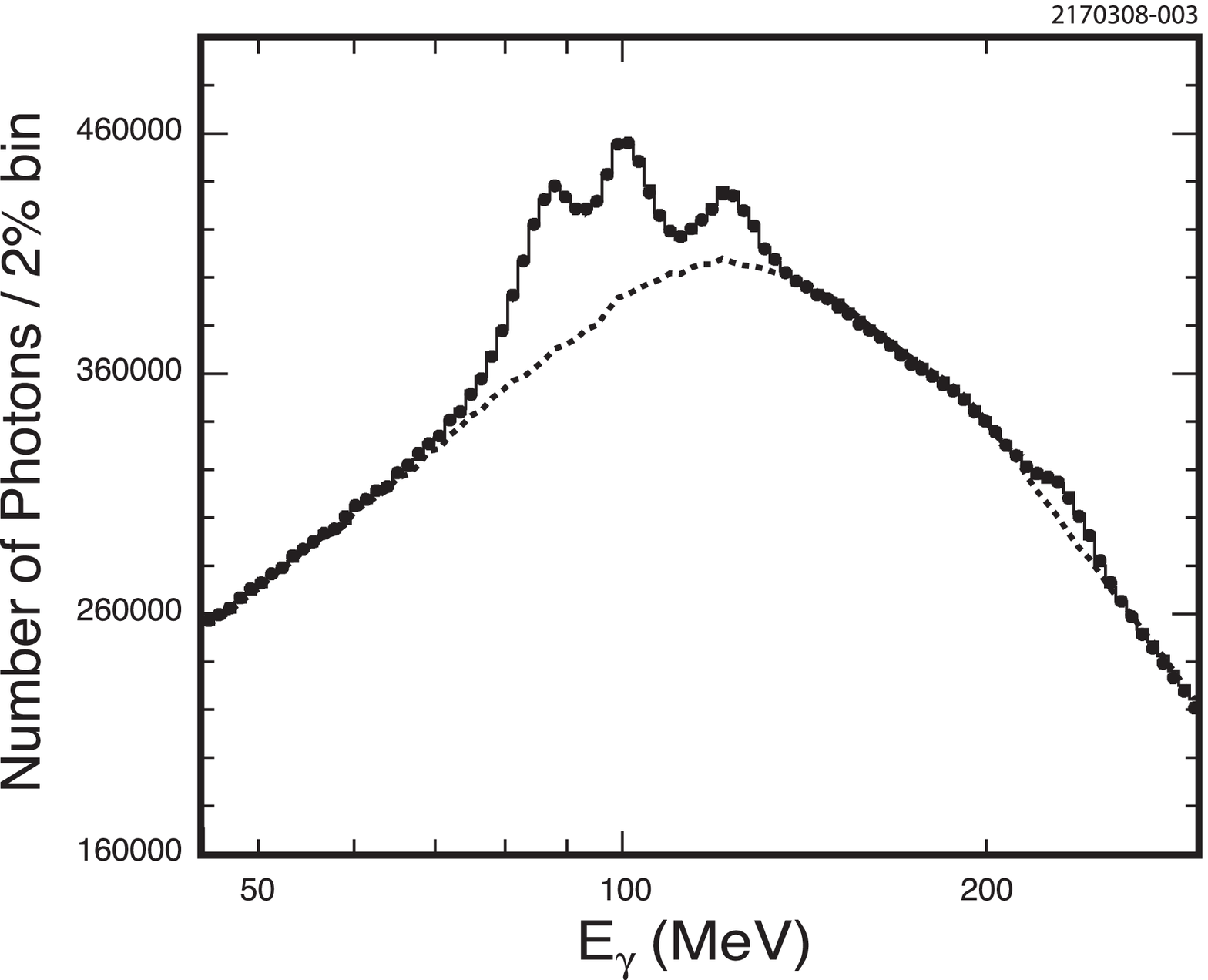}
}
\caption{\label{fig:Y23Sincl} 
Fits to the $\Upsilon(2S)$ (left) and $\Upsilon(3S)$ (right) 
inclusive photon energy spectra. 
The data are shown as dots; the fits are shown as the histograms; 
the dashed lines represents the total fitted background.  
Note the suppressed zero on the vertical axis.  
Nominal photon peak locations for transitions to the $\chi_{bJ}(1P)$ 
(on the left) are 111, 130, 164 MeV/$c^2$ (for $J = 2,1,0$, respectively) 
and for transitions to the $\chi_{bJ}(2P)$ (on the right) are 
87, 100, 123 MeV/$c^2$ (for $J = 2,1,0$, respectively).}  
\end{figure}

We first measure the total number of $\chi_{bJ}$ tagged with an observed 
transition photon by fitting the inclusive $E_{\gamma}$ spectrum.  
Photon peaks from inclusive $\Upsilon(2S) \to \gamma\chi_{bJ}(1P)$ and 
$\Upsilon(3S) \to \gamma\chi_{bJ}(2P)$ transitions are evident 
in Fig.~\ref{fig:Y23Sincl}.  

We use $\Upsilon(1S)$ resonance and $\Upsilon(nS)$ off-resonance data 
to model the photon background in the $E_{\gamma}$ 
spectra \cite{mura}.  The off-resonance data are observed to have 
indistinguishable spectra in our energy region and thus the three samples 
are combined to increase statistics.  
The $\Upsilon(1S)$ on-resonance and $\Upsilon(nS)$ off-resonance 
shapes are also quite similar, and we initially fit with two 
independent normalizations to peak-free regions of the photon 
energy spectrum.  The regions are defined by 
$3.50 <$ ln($E_\gamma$ [MeV]) $< 3.70$  ($33$ MeV $< E_\gamma < 40$ MeV) 
and 
$5.55 <$ ln($E_\gamma$ [MeV]) $< 5.70$ ($257$ MeV $< E_\gamma < 299$ MeV) 
and the fit results are used to then fix the relative normalization 
of these on- and off-resonance samples for subsequent signal fits.  

When fitting the full photon energy spectra to extract signal yields, 
only one overall normalization parameter for the background is varied.  
We find, however, that the fit quality is acceptable only 
after the inclusion of first- ($1P$) or second-order ($2P$) polynomials 
to allow small smooth adjustments of the background shape.  
The fit also includes signal contributions from 
the three dominant E1 transitions, $\Upsilon(2S) \to \gamma\chi_{bJ}(1P)$ 
or $\Upsilon(3S) \to \gamma\chi_{bJ}(2P)$, as appropriate.  
The $\chi_{bJ}(1P)$ and $\chi_{bJ}(2P)$ signal peaks are described by 
a so-called Crystal Ball line shape \cite{cbline} 
with fixed asymmetry parameters, $\alpha$ and $n$.  
This line shape is a Gaussian, described by a peak energy $E_p$ 
and resolution $\sigma_E$, matched with the constant $c$ onto an 
asymmetric low energy tail, $1/(E_p - E + c)^n$, 
at an energy $E_p - \alpha \sigma_e$.  
We obtain $E_p$ from published results \cite{pdg} and use 
the values $\alpha = 0.84$ and $n = 25.8$.  
The values of $sigma_E/E$ depend on $E$, varying from 
5.4\% to 3.9\% as the energy of the six transition lines increases.  
This $E$ dependence is determined from Monte-Carlo studies, but 
the overall scale of the resolution is adjusted based on fits to data.  
In addition to the dominant $\Upsilon(3S) \to \gamma\chi_{bJ}(2P)$ 
transitions,  the fit to the $\Upsilon(3S)$ spectrum includes the lines 
due to $\chi_{bJ}(2P)\to \gamma\Upsilon(2S)$ cascades.  
The fit results are displayed with the data in 
Fig.~\ref{fig:Y23Sincl} and tabulated in 
Tables~\ref{tab:t1} and \ref{tab:t2}.  

\begin{figure}
\parbox{0.48\textwidth}
{
 \includegraphics*[width=2.7in,height=2.5in]{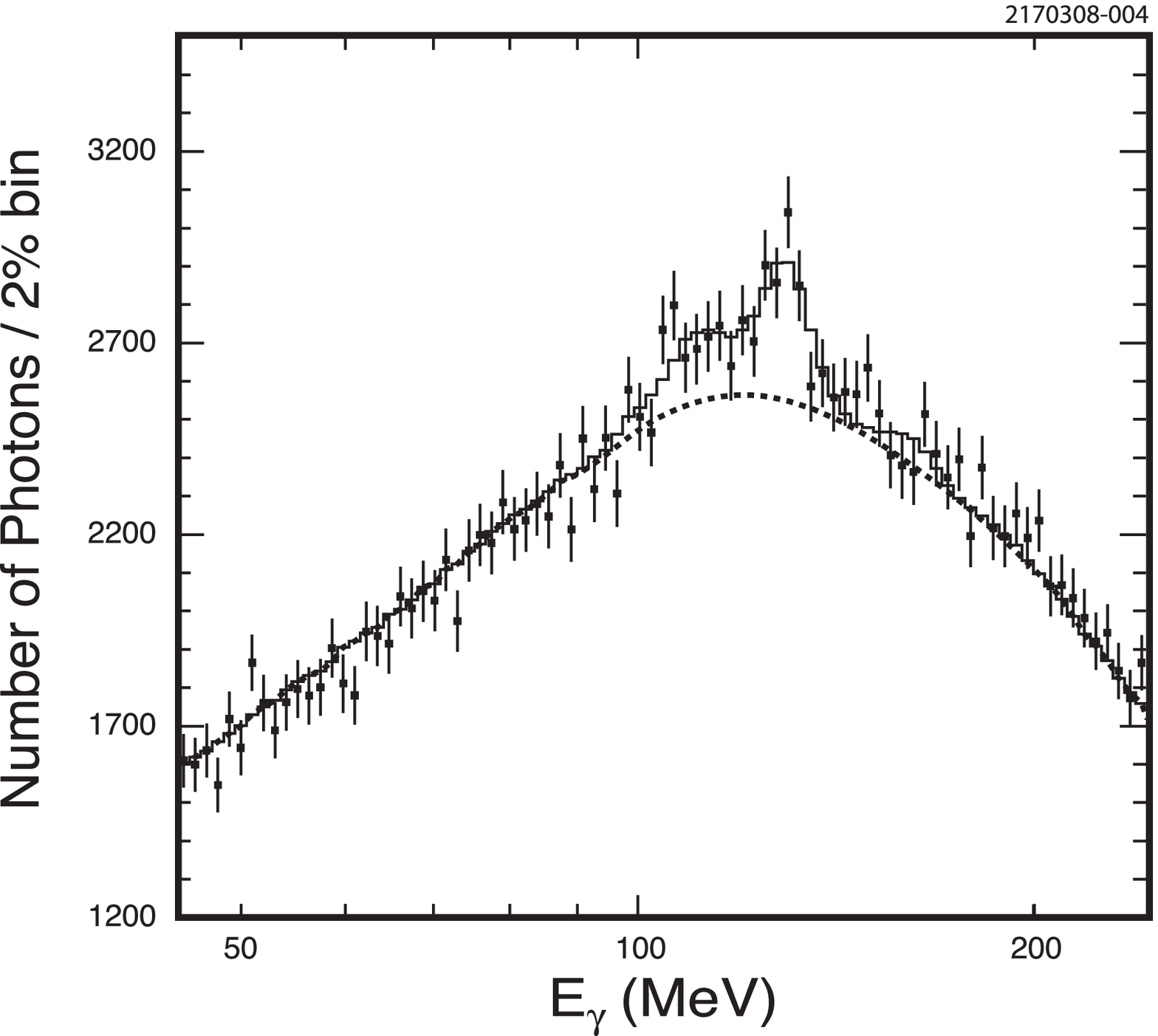}
}
\parbox{0.48\textwidth}
{
 \includegraphics*[width=2.7in,height=2.5in]{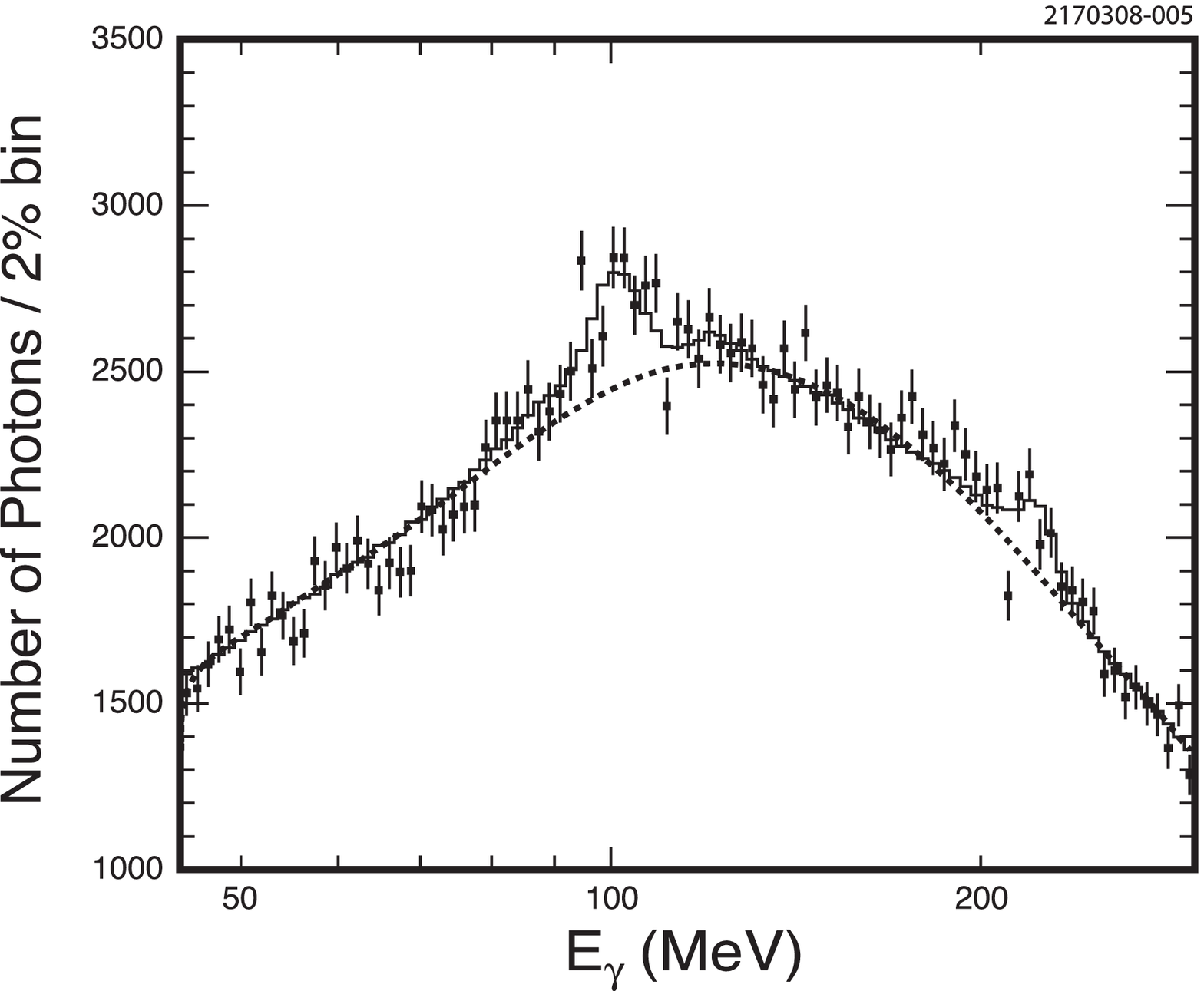}
}
\caption{\label{fig:Y23SwD0} 
Fits to the $\Upsilon(2S)$ (left) and $\Upsilon(3S)$ (right) 
photon energy spectrum obtained for events with $D^0$ mesons. 
The data are shown as dots; the fits are shown as histograms; 
the dashed lines represents the total fitted background.}
\end{figure}

\begin{figure}
\parbox{0.48\textwidth}
{
 \includegraphics*[width=2.7in,height=2.5in]{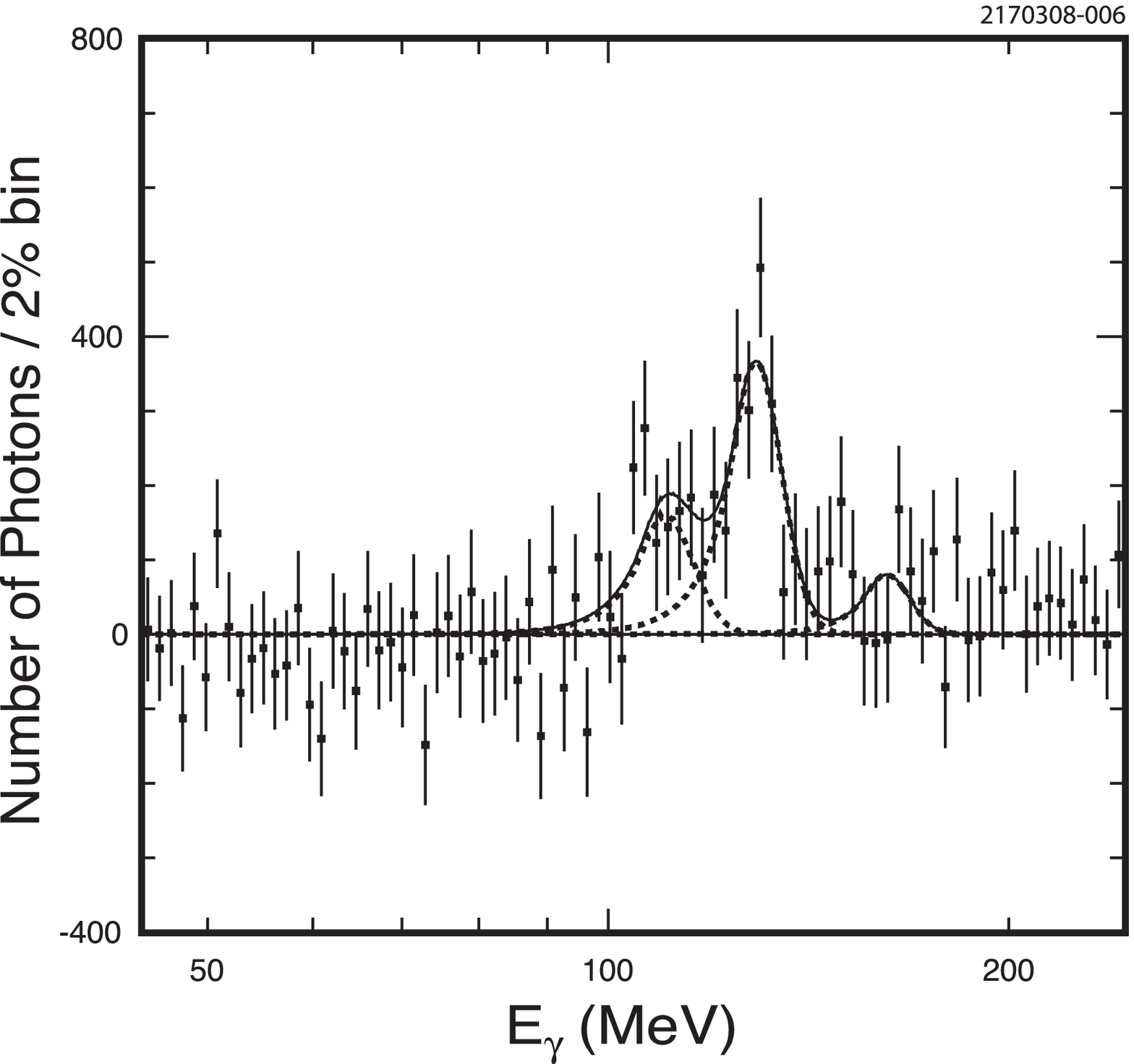}
}
\parbox{0.48\textwidth}
{
 \includegraphics*[width=2.7in,height=2.5in]{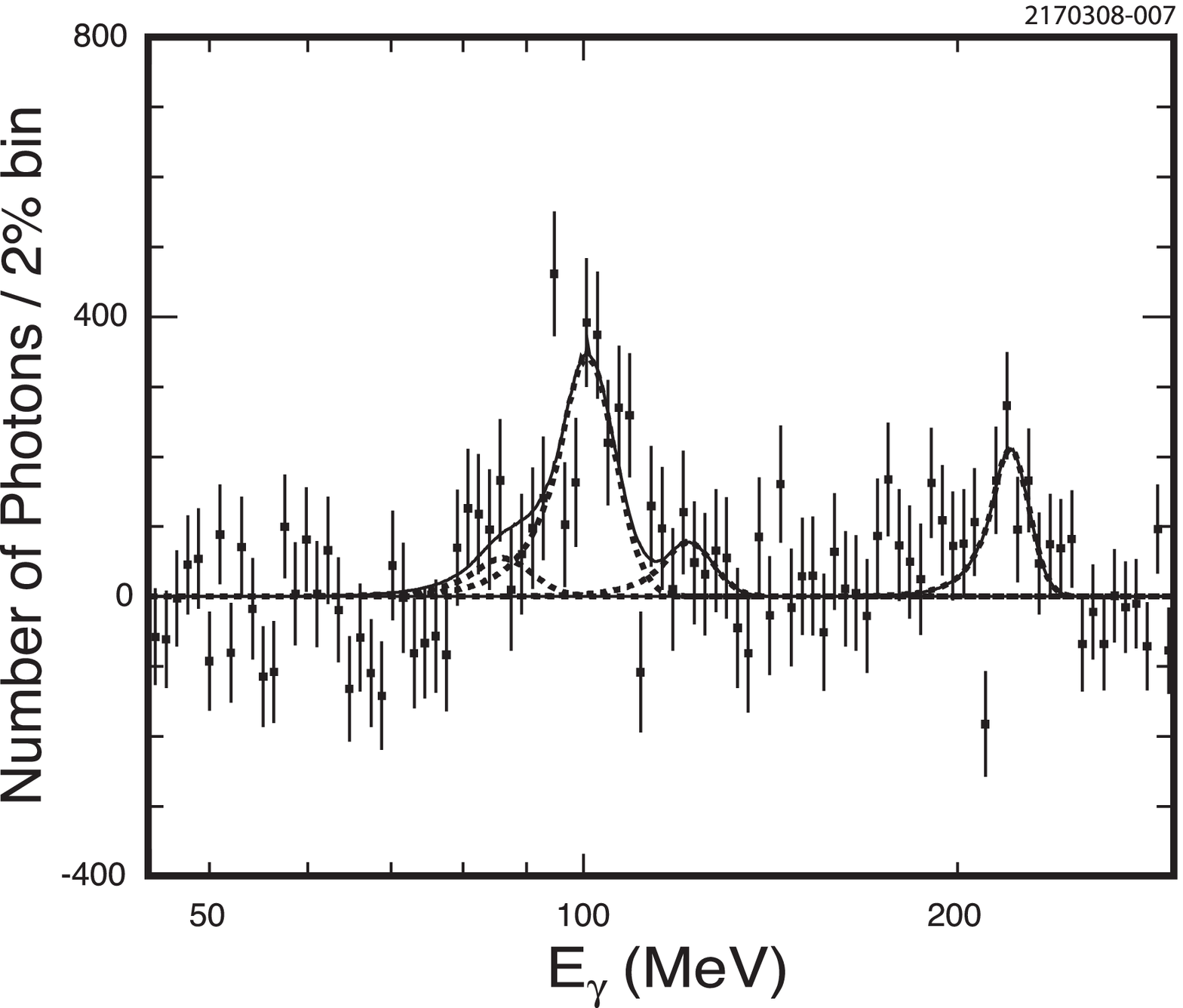}
}
\caption{\label{fig:Y23SwD0bs} 
Energy spectrum for background-subtracted 
$\Upsilon(2S) \to \gamma\chi_{bJ}(1P)$ (left) and 
$\Upsilon(3S) \to \gamma\chi_{bJ}(2P)$ (left) 
photon lines  obtained for events with $D^0$ mesons. 
The data are shown as dots; the fit is shown as the solid line.  
Individual contributions from the signal 
$\Upsilon(mS) \to \gamma\chi_{bJ}(nP)$ lines are shown as dashed-line peaks.}
\end{figure}

Photon energy spectra for events with $D^0$ mesons are obtained 
by subtracting the ln($E_\gamma$ [MeV]) spectra associated with the 
$Kn\pi$ ($n=1,2,3$) $D^0$ sidebands from the $D^0$ signal region.  
The ln($E_\gamma$ [MeV]) distributions and the fits for 
the $\Upsilon(2S)$ and $\Upsilon(3S)$ data 
are presented in Fig.~\ref{fig:Y23SwD0}.  
The $J=1$ lines are the most pronounced.  
Photon background shapes for these spectra are the same as for 
the $\Upsilon(2S)$ and $\Upsilon(3S)$ inclusive photon analysis, 
except that an acceptable fit quality is obtained without 
the addition of low-order polynomials, and they are omitted.  
The background-subtracted photon spectra are presented in 
Fig.~\ref{fig:Y23SwD0bs} and fit results are tabulated in 
Tables~\ref{tab:t1} and \ref{tab:t2}.  

\section{\boldmath Measurement of 
$\chi_{bJ}\rightarrow D^0 X$ 
(${\lowercase{p}}_{D^0} >  2.5$ G{\lowercase{e}}V/$c$) 
Rates}

The yields of events with $\chi_{bJ}$ and $D^0$ mesons 
($D^0 \to K^-\pi^+, K^-\pi^+\pi^0, K^-\pi^-\pi^+\pi^+$) 
include non-direct $\chi_{bJ}$ decays which must be subtracted.    
Non-direct $\chi_{bJ}(1P)$ decays to $D^0X$ include 
$\Upsilon(2S) \to \gamma\chi_{bJ}(1P); \chi_{bJ}(1P) \to \gamma\Upsilon(1S)$ 
decays where $D^0$ mesons are then produced in 
$\Upsilon(1S)$ annihilation into $ggg$, $gg\gamma$, and $\gamma$.  

Non-direct $\chi_{bJ}(2P)$ decays to $D^0X$ 
similarly include production of bottomonium states which 
in turn may decay to $D^0 X$.  
Known processes include $\Upsilon(1S)$ produced via   
$\Upsilon(3S) \to \gamma\chi_{bJ}(2P)$ followed by 
\begin{itemize}
\item $\chi_{bJ}(2P) \to (\gamma,\omega)\Upsilon(1S)$
\item $\chi_{bJ}(2P) \to \gamma\Upsilon(2S); 
       \Upsilon(2S) \to (\pi\pi, \pi^0,\eta)\Upsilon(1S)$ 
\item $\chi_{bJ}(2P) \to \gamma\Upsilon(2S); 
       \Upsilon(2S) \to \gamma\chi_{bJ}(1P); 
       \chi_{bJ}(1P) \to \gamma\Upsilon(1S)$ 
\item $\chi_{bJ}(2P) \to \pi\pi\chi_{bJ}(1P); 
       \chi_{bJ}(1P) \to \gamma\Upsilon(1S)$
\end{itemize}
and $\chi_{bJ}(1P)$ produced via $\Upsilon(3S) \to \gamma\chi_{bJ}(2P)$  
followed by
\begin{itemize}
\item $\chi_{bJ}(2P) \to \pi\pi\chi_{bJ}(1P)$
\item $\chi_{bJ}(2P) \to \gamma\Upsilon(2S); 
       \Upsilon(2S) \to \gamma\chi_{bJ}(1P)$
\end{itemize}
and $\Upsilon(2S)$ from $\Upsilon(3S) \to \gamma\chi_{bJ}(2P); 
\chi_{bJ}(2P) \to \gamma\Upsilon(2S)$.  

Yields for events with $D^0$ mesons from direct $\chi_{bJ}(1P)$ decays 
are calculated by correcting raw yields from the $\Upsilon(2S)$ 
data with a non-direct rate determined using known branching fractions 
\cite{pdg} and an $\Upsilon(1S) \to (ggg, gg\gamma,\gamma) 
\to D^0X$ rate for $p_{D^0} > 2.5$ GeV/$c$ 
of $2.60 \pm 0.50\%$~\cite{mw}.  
We estimate the numbers of these non-direct events as 
$16 \pm 9, 191 \pm 58$, and $125 \pm 34$ 
for $J = 0, 1$, and $2$, respectively.  
Corresponding estimates of the non-direct backgrounds for 
$\chi_{bJ}(2P) \to D^0 X$ in the $\Upsilon(3S)$ data are 
$53 \pm 24, 392 \pm 70$, and $311 \pm 50$ for $J = 0, 1$, and $2$, 
respectively.  
We account for the fact that prompt production of $D^0 X$ from 
$\Upsilon(2S)$ differs from that from $\Upsilon(1S)$ due to the 
different mixture of decays mediated by $ggg,gg\gamma$, and $\gamma$.  

Yields for inclusive $\chi_{bJ}$ production, total $\chi_{bJ}$ 
with $D^0$ mesons, and $\chi_{bJ}$ with directly-produced $D^0$ mesons, 
are summarized in Tables~\ref{tab:t1} and \ref{tab:t2}.  
In addition, we list a correction due to a small observed 
curvature in the $Kn\pi$ mass spectra leading to a small 
residual background of true photons and fake $D^0$ mesons, since 
our sideband subtraction assumes a flat background.  

The direct $\chi_{bJ}$ yields, $N_{\chi_{bJ}}^{D^0,dir}$, 
from $N_\Upsilon$ initial $\Upsilon$ produced are: 
$$ N_{\chi_{bJ}}^{D^0,dir} = N_\Upsilon \,
                       \epsilon_\gamma \, 
                       {\cal B}(\Upsilon \to \gamma \chi_{bJ}) \,
                       {\cal B}(\chi_{bJ} \to gg, q\bar{q}g \to D^0 \,\, X) \, 
                        \sum \epsilon_i {\cal B}_i(D^0),$$ 
where $\epsilon_\gamma$ is the $\gamma$ detection efficiency and 
the last factor $\sum \epsilon_i {\cal B}_i(D^0)$ is a sum 
over the three $Kn\pi$ decay modes of the $D^0$.  
The observed number of inclusive $\chi_{bJ}$ decays is given by 
$$
 N_{\chi_{bJ}}^{\mathrm{Incl}} = N_\Upsilon \, \epsilon_\gamma \, 
                       {\cal B}(\Upsilon \to \gamma \chi_{bJ}).
$$
Our main results, the branching fractions 
${\cal B}(\chi_{bJ} \to gg, q\bar{q}g \to D^0 X)$, 
are obtained from the two previous equations as:  
$$
{\cal B}(\chi_{bJ} \to gg, q\bar{q}g \to D^0 X) = 
  \frac{ N_{\chi_{bJ}}^{D^0} }
       { N_{\chi_{bJ}}^{\mathrm{Incl}} \, \sum \epsilon_i {\cal B}_i(D^0) }, 
$$
where the photon efficiency, $\epsilon_\gamma$, and sample 
size, $N_\Upsilon$, both cancel.  
For determination of the $D^0$ detection efficiencies, Monte-Carlo 
simulation of continuum $c\bar{c}$ events (based on Jetset 7 \cite{jetset})
were used, since this sample is expected to 
approximate the jet-like events from the $\chi_{bJ}\to c\bar{c}g$ 
decays.  We find that the efficiency is consistent with being independent 
of momentum in the $p_{D^0} > 2.5$~GeV/$c$ range.  

Based on detailed comparisons of particle identification 
in our data and Monte-Carlo simulations, we conclude that small efficiency 
corrections are needed.  
The $Kn\pi$ modes receive adjustments of $f_K  f_\pi^n$, where 
$f_K   = 0.95 (0.99)$ and 
$f_\pi = 0.99 (1.01)$ for $\Upsilon(2S)$ ($\Upsilon(3S)$) data.  
The $\chi_{bJ} \to D^0 X$ decay rates for $p_{D^0} >2.5$ GeV/$c$ 
are presented in Tables~\ref{tab:t1} and \ref{tab:t2}.  

\begin{table}[!ht]
\caption{\label{tab:t1} $\Upsilon(2S) \to \gamma\chi_{bJ}(1P)$ ($J = 0, 1, 2$) 
         transition yields and $\chi_{b} \to gg, q\bar{q}g \to D^0 X$ rates, 
         for $p_{D^0} > 2.5$ GeV/$c$.  Errors shown are statistical only.}
\renewcommand{\arraystretch}{1.25}
\begin{center}
\begin{tabular}{cccc}
\hline \hline
Final state&$\chi_{b0}(1P)$ &$\chi_{b1}(1P)$ & $\chi_{b2}(1P)$\\
 \hline
$N_{\chi_{bJ}}^{\mathrm{Incl}}$              
 & $166860 \pm  5988$ & $363825 \pm 6793$ & $379457 \pm 7243$ \\
$N_{\chi_{bJ}}^{D^0}$ (raw)   
 & $   501 \pm   303$ & $  2561 \pm  346$ & $  1207 \pm  360$ \\
$D^0$ sideband correction                
 & $    11 \pm     5$ & $    60 \pm    6$ & $    57 \pm    7$ \\
non-direct $D^0$                    
 & $    16 \pm     9$  & $  191 \pm   58$ & $   125 \pm   34$ \\
$N_{\chi_{bJ}}^{D^0,dir}$ (direct)
 & $   474 \pm   303$  & $ 2310 \pm  351$ & $  1025 \pm  362$ \\
${\bf \cal B}${\bf \boldmath ($\chi_{bJ}(1P) \to gg, q\bar{q}g \to D^0 X$)} 
 & {\bf \boldmath $ 5.63 \pm 3.61 \%$}
 & {\bf \boldmath $12.59 \pm 1.94 \%$}
 & {\bf \boldmath $ 5.36 \pm 1.90 \%$} \\ 
\hline\hline

\end{tabular}
\end{center}
\end{table}

\begin{table}[!ht]
\caption{\label{tab:t2} $\Upsilon(3S) \to \gamma\chi_{bJ}(2P)$ ($J = 0, 1, 2$) 
         transition yields and $\chi_{b} \to gg, q\bar{q}g \to D^0 X$ rates, 
         for $p_{D^0} > 2.5$ GeV/$c$.  Errors shown are statistical only.}
\renewcommand{\arraystretch}{1.25}
\begin{center}
\begin{tabular}{cccc}
\hline \hline
Final state & $\chi_{b0}(2P)$ & $\chi_{b1}(2P)$ & $\chi_{b2}(2P)$ \\ 
 \hline
$N_{\chi_{bJ}}^{\mathrm{Incl}}$              
 & $219773 \pm  5201$ & $491818 \pm 5197$ & $524549 \pm 5628$ \\
$N_{\chi_{bJ}}^{D^0}$ (raw)   
 & $   565 \pm   341$ & $  2757 \pm  366$ & $   477 \pm  370$ \\
$D^0$ sideband correction                
 & $    39 \pm     7$ & $   122 \pm    7$ & $   122 \pm    7$ \\
non-direct $D^0$                    
 & $    53 \pm    24$ & $   392 \pm   70$  & $  311 \pm   50$ \\
$N_{\chi_{bJ}}^{D^0,dir}$ (direct)
 & $   473 \pm   342$  & $  2243 \pm 373$  & $   44 \pm  373$ \\
${\bf \cal B}${\bf \boldmath ($\chi_{bJ}(2P) \to gg, q\bar{q}g \to D^0 X$)} 
 & {\bf \boldmath $4.13 \pm 3.00 \%$}
 & {\bf \boldmath $8.75 \pm 1.47 \%$}
 & {\bf \boldmath $0.16 \pm 1.37 \%$} \\ 
\hline\hline

\end{tabular}
\end{center}
\end{table}

\section{Systematic Uncertainties on the Branching Fractions}

Systematic uncertainties on the six measured branching fractions 
are primarily of two types.  The first are uncertainties in $D^0$ 
reconstruction; these affect each of the six $\chi_{bJ}$ states equally 
and are summarized in Table \ref{tab:syst_br1}.  
The next are uncertainties related to our photon yields, 
both in terms of efficiencies and yield extractions.  
These often differ for the six $\chi_{bJ}$ states, 
and are summarized in Table \ref{tab:syst_br2}.  
In the remainder of this section we detail the sources of 
the uncertainty estimates presented in the aforementioned Tables.  


The first three entries of Table \ref{tab:syst_br1} involve 
efficiencies for track-finding, $\pi^0$ reconstruction, 
and particle identification algorithms.  
Since the composition of the three $D^0$ final states differ, 
we take a linear weighting of the uncertainties across $D^0$ modes.  
The weights used are 
$w_i=\epsilon_i{\cal B}_i/\sum_j \epsilon_j{\cal B}_j$, 
yielding 0.25, 0.34, and 0.41 for $D^0\to K^-\pi^+$, 
$D^0\to K^-\pi^+\pi^0$, and $D^0\to K^-\pi^-\pi^+\pi^+$, respectively.  

The systematic uncertainty in track-finding is obtained by studies 
of the difference between data and Monte-Carlo simulation.  
We assign a 1.5\% uncertainty per track, which gives a net uncertainty 
of 4.2\% after weighting across $D^0$ decay modes.  

We assess the uncertainty in $\pi^0$-finding at 5\% per $\pi^0$. 
Taking into account the weight of the $D^0 \to K^-\pi^-\pi^0$ mode, 
the net $\pi^0$-finding systematic uncertainty is 1.7\%.  

Systematic uncertainties in kaon and pion identification are obtained 
by comparing data and Monte-Carlo efficiencies.  
We obtain 2\% (1\%) uncertainties per $K$ ($\pi$) which 
yield a net 4.0\% systematic uncertainty, averaged over $D^0$ modes.  

The systematic uncertainty on the $D^0 \to K^-\pi^+\pi^0$ 
efficiency due to selection on the Dalitz region is obtained 
by comparing the inclusive yield changes in data compared to 
Monte-Carlo simulations as the selection efficiency is varied.  
As a result of this study, and accounting for the fraction of 
$D^0$ candidates found via this decay mode, we assign 1.0\% as our 
total Dalitz region selection uncertainty.  

For evaluation of systematic uncertainties related to the 
$D^0$ momentum requirement, the $p_{D^0}$ requirement was varied.  
Events were selected for three values of the $D^0$ momentum 
requirement ($> 2.2, > 2.5$, and  $> 2.8$ GeV/$c$).  
We assign a 1.7\% branching fraction uncertainty due to 
this source.  

To study possible effects of the event shape and environment 
on the $D^0$ detection efficiency, different models of signal 
Monte-Carlo and continuum Monte-Carlo events are analyzed.  
Results indicate a 3.0\% uncertainty of the efficiency for 
the event-shape changes explored.  

Systematic uncertainties related to the definition of the 
$D^0$ signal and sideband regions are obtained by 
varying the corresponding mass windows.  
This also includes uncertainty due to a nonlinear background 
shape under the $D^0$ signal.  
The total systematic uncertainty is determined to be 2.5\%.  

The total uncertainty in the $D^0$ efficiency is $7.5\%$ 
for each $\chi_{bJ}$ state, as noted in Table \ref{tab:syst_br1}.  
We now turn to the photon-related systematic uncertainties 
presented in Table~\ref{tab:syst_br2}.  

\begin{table}[htb]
 \caption{Relative systematic uncertainties on measured branching fractions 
          from sources affecting the $D^0$ efficiency.}
 \begin{center}
 \centering
 \begin{tabular}{lc} 
 \hline\hline  
Source & {Uncertainty (\%)}\\
\hline
Tracking: 1.5\%/track                          & 4.2 \\
$\pi^0$ efficiency: 5\%/$\pi^0$                & 1.7 \\
PID: $2\%/K^\pm$, $1\%/\pi^\pm$                 & 4.0 \\
$K\pi\pi^0$ Dalitz requirement                 & 1.0 \\
Momentum dependence                            & 1.7 \\
Decay model effects on $D^0$ efficiency        & 3.0 \\ 
Selection of events with a $D^0$               & 2.5 \\\hline
Total $D^0$-related systematic uncertainty     & 7.5 \\\hline\hline
 \end{tabular}
 \end{center}
\label{tab:syst_br1}
\end{table}


To verify that the photon efficiency largely cancels in our analysis, 
the difference of photon efficiencies between inclusive events and 
those with a $D^0$ candidate is studied using Monte-Carlo samples.  
We find that the relative photon efficiency difference between spherical 
$ggg$ events and jet-like $q\bar{q}$ events is about 6\%.  
In our case, we are concerned about the difference between 
generic $\chi_{bJ}$ events and those having a reconstructed $D^0$.  
Presumably the effect of this bias is smaller than that of the rather 
large overall event shape change between these two Monte-Carlo samples.  
We thus take 1/3 of the variation and assign a 2\% uncertainty 
for all six $\chi_{bJ}$ states.  

For estimation of line-shape fitting uncertainties we change the 
Crystal Ball line-shape parameters $\alpha$ and $n$ by $\pm 10\%$ 
from their nominal values.  
This range is chosen as appropriate based on changes in fit quality. 
We take the resulting branching fraction variations 
as systematic uncertainties, ranging from 0.1\% to 0.6\%.  

The nominal fitting ranges for photon energy distributions are 
$3.8 <$ ln($E_\gamma$ [MeV]) $< 5.5$ for $\Upsilon(2S)$ and  
$3.8 <$ ln($E_\gamma$ [MeV]) $< 5.7$ for $\Upsilon(3S)$.  
We vary the lower and upper limits of the fitting regions
from 3.50 to 3.70 and from 5.50 to 5.70 .   
Variations in our results suggest uncertainties from 0.3\% to 0.6\%.  

As mentioned above, the photon background shape consists of two components: 
the resonant and off-resonance photon spectra used to estimate the background
shapes in the $\Upsilon(2S)$ and $\Upsilon(3S)$ photon energy distributions.   
We varied scaling factors for the photon background components and 
changed the $\Upsilon(1S)$ resonance and the $\Upsilon(2S)$, $\Upsilon(3S)$ 
off-resonance contributions in the photon background shape. 
Also, in the fit of the $\Upsilon(2S)$ and $\Upsilon(3S)$ 
inclusive photon energy distributions, 
we used additional background components to obtain a better fit quality.  
First, second, and third order polynomials are tried as extra components 
in addition to the $\Upsilon(1S)$ on-resonance 
and the $\Upsilon(2S)$ and $\Upsilon(3S)$ off-resonance background shapes.  
We estimate systematic uncertainties due to such choices at levels 
ranging from 0.5\% to 1.6\%.

Our nominal fit uses logarithmic binning of energy ln($E_\gamma$ [MeV]).  
We changed the logarithmic energy scale to linear binning, 
with 1 MeV energy bins.  
The photon background shape was left unchanged.  
We assign from 0.2\% to 1.7\% uncertainties on our branching fractions 
based on the stability of our results.  

The $\Upsilon(3S)$ photon energy spectrum includes 
$\Upsilon(2S) \to \gamma\chi_{bJ}(1P)$ transition lines at similar energies.  
To estimate systematic uncertainties 
on the ${\cal B}(\chi_{bJ}(2P)\to D^0X)$, 
we include these lines in the fit to the $\Upsilon(3S)$ inclusive 
photon spectrum and the photon spectrum for events with $D^0$ mesons.  
Estimated systematic uncertainties varied from 0.2\% to 1.5\%.  
 
In Table \ref{tab:syst_br2}, we summarize the systematic uncertainties 
associated with $\gamma$ detection and fitting 
for each of the six $\chi_{bJ}$ lines.  
Note that these uncertainties apply to the raw yields,   
before any subtractions are made.  

\begin{table}[htb]
 \caption{Relative systematic uncertainties on measured branching fractions 
          due to sources related to the $E_\gamma$ distributions.} 
 \begin{center}
 
 \renewcommand{\arraystretch}{1.25}
 \begin{tabular}{lcccccc} 
 \hline\hline 
 &\multicolumn{6}{c}{Uncertainty (\%)}\\
\cline{2-7}
Source & $\chi_{b0}(1P)$\, & $\chi_{b1}(1P)$\, & $\chi_{b2}(1P)$\, 
       & $\chi_{b0}(2P)$\, & $\chi_{b1}(2P)$\, & $\chi_{b2}(2P)$ \\
\hline
$\gamma$ efficiency cancellation      & 2.0 & 2.0 & 2.0 & 2.0 & 2.0 & 2.0  \\
Line-shape fitting        & 0.6 & 0.1 & 0.4 & 0.5 & 0.1 & 0.5  \\
Fitting range             & 0.5 & 0.3 & 0.4 & 0.6 & 0.3 & 0.5  \\
Background shape          & 1.4 & 0.6 & 0.9 & 1.6 & 0.5 & 0.9  \\
$\gamma$ energy binning   & 1.4 & 0.2 & 0.5 & 1.7 & 0.3 & 0.6  \\
$\Upsilon(2S) \to \gamma\chi_{bJ}(1P)$ lines 
                          & --  & --  & --  & 1.5 & 0.3 & 0.2  \\
\hline
Total $\gamma$ systematic uncertainty & 2.9 & 2.1 & 2.3 & 3.5 & 2.1 & 2.4  \\
\hline\hline
 \end{tabular}
 \end{center}
\label{tab:syst_br2}
\end{table}

We also performed several simple cross-checks to investigate 
the stability and consistency of our results.  
These included splitting the datasets into two subsets, 
varying selection criteria, and comparing yields in individual 
$D^0$ decay modes.  All of these tests produced consistent results.

Our final results for $p_{D^0} > 2.5$ GeV/$c$ are 
given in Table \ref{tab:rawbr}.  
Upper limits are given for modes without significant signals, 
but central values for those modes will be needed for fits 
later. 

\begin{table}[htb]
 \caption{Summary of measured branching fractions (or upper limits) for 
   ${\cal B}$($\chi_{bJ}(nP)\to gg, q\bar{q}g \to D^0X$) 
   with the requirement that $p_{D^0} > 2.5$ GeV/$c$.  
   The uncertainties are statistical and systematic, respectively.}
 \begin{center}
  \begin{tabular}{ccc} 
 \hline\hline 
   State & ${\cal B}$($\chi_{bJ}(nP)\to gg, q\bar{q}g \to D^0X$) (\%) 
         & 90\% CL UL (\%) \\  
  \hline
$\chi_{b0}(1P)$ & $ 5.6 \pm 3.6 \pm 0.5$ & $< 10.4$ \\
$\chi_{b1}(1P)$ & $12.6 \pm 1.9 \pm 1.1$ &   \\
$\chi_{b2}(1P)$ & $ 5.4 \pm 1.9 \pm 0.5$ & $<  7.9$ \\
$\chi_{b0}(2P)$ & $ 4.1 \pm 3.0 \pm 0.4$ & $<  8.2$ \\
$\chi_{b1}(2P)$ & $ 8.8 \pm 1.5 \pm 0.8$ &    \\
$\chi_{b2}(2P)$ & $ 0.2 \pm 1.4 \pm 0.1$ & $<  2.4$ \\

\hline\hline
 \end{tabular}
 \end{center}
\label{tab:rawbr}
\end{table}

\section{Interpretation}

We observe significant production of $D^0$ mesons from both the 
$\chi_{b1}(1P)$ and $\chi_{b1}(2P)$ states.  There is evidence of 
a signal for $\chi_{b2}(1P)$, while data for the other three states are 
inconclusive.  For each triplet, we observe the largest branching 
fraction for the $J=1$ states, as expected.  

The NRQCD calculation mentioned earlier \cite{Bodwin07} makes 
predictions for the total $c\bar{c}X$ production rate, $R^{(c)}_J$, 
as a function of one non-perturbative parameter, $\rho_8$, 
per $\chi_{bJ}$ triplet.  
We would like to convert our measurement of the inclusive $D^0 X$ rate, 
with a minimum momentum requirement, into an experimental 
value for $R^{(c)}_J$.  
However, this conversion {\it also} depends on $\rho_8$, since this 
parameter affects the momentum spectrum of the $D^0$ mesons and 
hence the efficiency of our minimum momentum requirement.  
We use six branching fraction results to determine two best-fit 
values of $\rho_8$ (one per triplet).  Our experimental results 
for $R^{(c)}_J$ are based on these best-fit values and clearly depend 
on our use of the NRQCD calculation.  

We first discuss the details of how to relate our measurements 
to the inclusive $c\bar{c}X$ rate and then present 
our extraction of the $\rho_8$ parameter and experimental values of 
$R^{(c)}_J$.  
Three factors will combine to cause our extracted $R^{(c)}_J$ to be larger 
than the directly-measured branching fractions in Table \ref{tab:rawbr}.  
We only see some of the $D^0$ spectrum, not all charm appears as $D^0$, 
and $R^{(c)}_J$ is normalized to the number of $\chi_{bJ}$ 
that decay via annihilation, not the total number produced.  
Only one factor works in the other direction: 
$R^{(c)}_J$ measures $c\bar{c}X$ production, and either charm 
quark may form a $D^0$.  

Suppressing the  $\chi_{bJ}(nP)$ radial quantum numbers for simplicity, 
we have: 
$$ 
  R^{(c)}_J = 
    \frac{ {\cal B}(\chi_{bJ} \to gg, q\bar{q}g \to c\bar{c}X) }
         { {\cal B}(\chi_{bJ} \to gg, q\bar{q}g) } = 
  \frac{ {\cal B}(\chi_{bJ} \to gg, q\bar{q}g \to D^0X,\,
                                p_{D^0} > 2.5\,{\rm GeV/}c) }
       { f_{2.5} \, f_{D^0} \, {\cal B}(\chi_{bJ} \to gg, q\bar{q}g) }, 
$$
where the right-hand side contains our directly-measured branching fraction 
with three additional factors which we now explain.  

First, we must divide by ${\cal B}(\chi \to gg, q\bar{q}g)$ such that 
the final branching fraction is normalized to only $gg, q\bar{q}g$ 
decays of the $\chi_{bJ}$ since this is the normalization used 
for the theoretical prediction.  
These branching fractions are calculated as $1 - \sum_i {\cal B}_k$, 
where the sum extends over all known transitions of a given 
$\chi_{bJ}$ to other bottomonium states \cite{pdg}.  

Next, we divide by $f_{2.5}$, the fraction of the $D^0$ spectrum 
expected to be above our 2.5~GeV/$c$ $D^0$ momentum requirement.  
This is obtained from the results of Ref. \cite{Bodwin07}, 
and it depends on the value of $\rho_8$ and knowledge of 
the charm fragmentation function \cite{BelleFrag}.  

Finally, we must divide by the number of $D^0$ mesons expected per 
$c\bar{c}X$ event: $f_{D^0} = 1.11 \pm 0.08$.  
This number is itself the product of four factors.  
The first is a factor of two to account for the two quarks, 
each of which may form a $D^0$.  
The next two factors account for all seven weakly-decaying $C=1$ states 
$D^0$, $D^+$, $D_s$, $\Lambda_c$, $\Xi_c^+$, $\Xi_c^0$, and $\Omega_c^0$, 
relative to the measured $D^0$ yields.  
The fraction of $D^0$ compared to the total of $D^0 + D^+ + D_s +\Lambda_c$, 
$N(D^0)/[N(D^0) + N(D^+) + N(D_s) + N(\Lambda_c)] = 0.574 \pm 0.041$, 
is obtained from $e^+ e^-$ fragmentation data \cite{BelleFrag}.  
An additional factor $0.98 \pm 0.01$ then accounts for 
the omitted $\Xi_c^+$, $\Xi_c^0$, and $\Omega_c^0$ states.  
This is estimated from the $\Lambda_c$ fraction of 
$N(\Lambda_c)/[N(D^0) + N(D^+) + N(D_s)] = (8.1 \pm 2.1) \%$ 
in \cite{BelleFrag} (with an added uncertainty 
from knowledge of ${\cal B}(\Lambda_c \to pK\pi)$), 
combined with a theoretical suppression of order $10\%$ due to the 
additional strange quark popping needed to form the omitted states.  
The fourth factor of $0.99 \pm 0.01$ accounts for charmonium states, 
which here include those states below open-flavor threshold 
at $\sqrt{s} = 2 M_{D^0}$: $J/\psi, \psi(2S), \eta_c, \eta_c(2P), 
\chi_{cJ}, h_c$.  
We estimate 
$N({\rm open}\,c)/[N({\rm open}\,c)+2N(c\bar{c})] 
 \simeq 1 - 2N(c\bar{c})/N({\rm open}\,c) 
 \simeq 1 - 2{\cal B}(c\bar{c}X \to {\rm charmonia})$ 
based on the production rate of $J/\psi$ in $e^+ e^-$ fragmentation 
\cite{BelleContOnia} and the branching fractions to charmonium 
in $\Upsilon(1S)$ decays \cite{pdg}; these processes show that charmonium 
is rare in both $\gamma$ and $ggg$ hadronization.    
We are not sensitive to errors at the 1\% level and choose 
a conservative uncertainty to accommodate unmeasured charmonium states.  
The various factors required for the six $\chi_{bJ}$ states 
are summarized in Table \ref{tab:interp}.  

\begin{table}[htb]
 \caption{Summary of factors used to relate our measured $D^0 X$ 
          branching fractions to $R^{(c)}_J$, 
          which measures the total $c\bar{c}X$ rate.  
          The values of $f_{2.5}$ are evaluated at the independently fitted 
          best values of $\rho_8$ for each triplet.} 
 \begin{center}
  \begin{tabular}{lcccccc} 
 \hline\hline 
Factor & $\chi_{b0}(1P)$ & $\chi_{b1}(1P)$ & $\chi_{b2}(1P)$ 
       & $\chi_{b0}(2P)$ & $\chi_{b1}(2P)$ & $\chi_{b2}(2P)$ \\
\hline
${\cal B}(\chi \to gg, q\bar{q}g)$ 
    & $0.97 \pm 0.03$ & $0.65 \pm 0.08$ & $0.78 \pm 0.04$ 
    & $0.93 \pm 0.07$ & $0.68 \pm 0.04$ & $0.75 \pm 0.03$ \\

$f_{2.5}$ 
    & $ 0.54 $ & $ 0.70 $ & $ 0.63 $ 
    & $ 0.45 $ & $ 0.46 $ & $ 0.47 $ \\
$f_{D^0}$ 
    & $ 1.11 \pm 0.08 $ & $ 1.11 \pm 0.08 $ & $ 1.11 \pm 0.08 $
    & $ 1.11 \pm 0.08 $ & $ 1.11 \pm 0.08 $ & $ 1.11 \pm 0.08 $\\
\hline
$1/(f_{D^0} f_{2.5} {\cal B})$ 
    & $1.70 \pm 0.13$ & $1.97 \pm 0.28$ & $1.83 \pm 0.16$ 
    & $2.15 \pm 0.23$ & $2.89 \pm 0.28$ & $2.56 \pm 0.21$ \\
\hline\hline
 \end{tabular}
 \end{center}
\label{tab:interp}
\end{table}

With these factors in hand, we fit our data for the 
$D^0 X$ branching fractions with $p_{D^0} >$ 2.5 GeV/$c$ to the 
NRQCD predictions \cite{Bodwin07} and extract $\rho_8$, 
the ratio of color-octet to color-singlet matrix elements, 
in $\chi_{bJ}$ decays.  
Recall that both $f_{2.5}$  and $R^{(c)}_j$ depend on $\rho_8$ 
and that $f_{2.5}$ depends on fragmentation functions.  
For each value of $\rho _8$, we may convert our directly measured 
branching fractions into extracted values for $R^{(c)}_J$ 
in the context of this NRQCD calculation (which includes the 
assumption that $e^+e^-$ charm fragmentation data is representative 
of our charm fragmentation).  
The best value of $\rho_8$ is obtained from a fit which finds the 
best agreement between the predicted and extracted $R^{(c)}_J$.  

We fit separate $\rho_8$ values for each triplet by minimizing a $\chi^2$ 
which has one term for each of the three states.  
Each term in the $\chi^2$ is formed from the square of the deviation 
of the predicted and extracted $R^{(c)}_J$ values, 
normalized by the errors on the extracted value.  
Note that {\it both} the predicted and extracted $R^{(c)}_J$ values 
depend on $\rho_8$.  
Correlated systematic uncertainties on the branching fractions 
are incorporated into the covariance matrix used to evaluate 
the $\chi^2$ in our fits.  We find, however, that results are insensitive 
to correlations due to the dominance of statistical errors.  
The best-fit values are 
$\rho_8(1P) = 0.160 ^{+0.071} _{-0.047}$ and 
$\rho_8(2P) = 0.074 ^{+0.010} _{-0.008}$ 
with $\chi^2(1P) = 0.40$ and $\chi^2(2P) = 4.71$, respectively, 
for $3-1$ degrees of freedom each.  
The errors are larger for the $1P$ states primarily due to the 
non-linear dependence of the branching fractions on $\rho_8$: 
for larger $\rho_8$, the branching fractions are less sensitive 
to changes in its value.  

It has been argued \cite{brambilla} that $\rho_8$ should be 
largely independent of radial quantum number.  
While we prefer not to assume such an equality, a joint fit to our 
branching fractions for both triplets obtains a best-fit common value of 
$\rho_8 = 0.086^{+0.009}_{-0.013}$, with $\chi^2 = 10.1$ for $6-1$ degrees 
of freedom.  

Table \ref{tab:fitbr} lists the best-fit branching fractions, $R^{(c)}_J$, 
extracted from our data along with the best-fit NRQCD values, based on 
fits with separate $\rho_8$ parameters for each $\chi_{bJ}$ triplet.  
We also show the original 1979 calculations \cite {Barb79} for comparison.  
The third uncertainty is due to uncertainties in the branching fractions 
used to obtain ${\cal B}(\chi \to gg, q\bar{q}g)$ 
and the fragmentation data used to obtain $f_{D^0}$ and $f_{2.5}$.  
No systematic uncertainty is included for the accuracy of the 
theoretical calculations or the assumption that the $e^+ e^-$ 
fragmentation data is a valid model for our charm fragmentation 
since we do not know how to quantify such effects.  
Thus, while our primary results for the inclusive $\chi_{bJ}$ 
branching fractions into $D^0 X$ with $p_{D^0} > 2.5$ GeV/c 
are model-independent, our results for $R^{(c)}_J$ are clearly 
model-dependent.  
 
\begin{table}[htb]
 \caption{Summary of extracted branching fractions (or upper limits) 
   for $R^{(c)}_J$.  
   NRQCD best-fit values use distinct $\rho_8$ values for each 
   $\chi_{bJ}$ triplet.  
   The original 1979 calculations \cite {Barb79} are also shown.  
   The uncertainties are statistical, our systematic, and external 
   systematic, respectively.}
 \begin{center}
  \begin{tabular}{ccccc} 
 \hline\hline 
   State & $R^{(c)}_J$ (\%) 
         & 90\% CL UL (\%) & NRQCD Best Fit (\%) 
         & Pred. from \cite {Barb79} (\%) \\  
  \hline
$\chi_{b0}(1P)$ &
    $9.6 \pm 6.2 \pm 0.8 \pm 0.8$ & $<17.9$ &  6.3 &  6 \\
$\chi_{b1}(1P)$ &
   $24.8 \pm 3.8 \pm 2.2 \pm 3.6$ &         & 23.7 & 25 \\
$\chi_{b2}(1P)$ &
   $ 9.8 \pm 3.5 \pm 0.9 \pm 0.9$ & $<14.6$ & 10.8 & 12 \\
$\chi_{b0}(2P)$ &
   $ 8.7 \pm 6.4 \pm 0.9 \pm 0.7$ & $<17.7$ &  4.9 &  6\\
$\chi_{b1}(2P)$ &
   $25.3 \pm 4.3 \pm 2.5 \pm 2.4$ &         & 22.1 & 25\\
$\chi_{b2}(2P)$ &
   $ 0.4 \pm 3.5 \pm 0.4 \pm 0.1$ &  $<6.1$ &  7.4 & 12\\
\hline\hline
 \end{tabular}
 \end{center}
\label{tab:fitbr}
\end{table}

\section{Conclusion}

We report first measurements of the branching fractions 
for $\chi_{bJ}(1P,2P) \to D^0 X$ with $p_{D^0} > 2.5$ GeV/$c$.  
Our results are used to infer the total production of charm 
in $\chi_{bJ}$ decays, $R^{(c)}_J$ in the context of a recent 
NRQCD calculation \cite{Bodwin07}.  The results are in agreement 
with this calculation, as well as the older calculations \cite{Barb79}.  
Notably, our $R^{(c)}_J$ values confirm that the largest branching 
fractions to charm correspond to the $J=1$ $\chi_{bJ}$ states.

\section{ACKNOWLEDGMENTS}

We thank the authors of Ref. \cite{Bodwin07} for providing convenient 
parameterizations of their results.  
We gratefully acknowledge the effort of the CESR staff
in providing us with excellent luminosity and running conditions.
D.~Cronin-Hennessy and A.~Ryd thank the A.P.~Sloan Foundation.
This work was supported by the National Science Foundation,
the U.S. Department of Energy,
the Natural Sciences and Engineering Research Council of Canada, and
the U.K. Science and Technology Facilities Council.


\begin{thebibliography}{99}

\bibitem{pdg} 
W.-M.~Yao {\it et~al.} (Particle Data Group), 
{J. Phys. G: Nucl. Part. Phys.} \textbf{33}, 1 (2006).

\bibitem{CLEOchihad} 
C.~Cawlfield {\it et~al.} (CLEO Collaboration), 
Phys. Rev. D \textbf{73}, 012003 (2006); 
D.~Cronin-Hennessy {\it et~al.} (CLEO Collaboration), 
Phys. Rev. Lett. \textbf{92}, 222002 (2004).  

\bibitem{Barb76} 
R.~Barbieri, R.~Gatta, and R.~K\"{o}gerler, 
{Phys. Lett. B} \textbf{60}, {183} (1976).  

\bibitem{ARGUS}
H. Albrecht {\it et al.} (ARGUS Collaboration), 
Z. Phys. C \textbf{55}, 25 (1992).  

\bibitem{Fritzsch78}
H.~Fritzsch and K.-H.~Streng, 
{Phys. Lett. B} \textbf{77}, 299 (1978).  

\bibitem{Barb79} 
R.~Barbieri, M.~Caffo, and E.~Remiddi, 
{Phys. Lett. B} \textbf{83}, 345 (1979).  

\bibitem{NRQCD92}
G.T.~Bodwin, E.~Braaten, and G.P.~Lepage, 
Phys. Rev. D \textbf{46}, R1914 (1992).  

\bibitem{NRQCD95}
G.T.~Bodwin, E.~Braaten, and G.P.~Lepage, 
Phys. Rev. D \textbf{51}, 1125 (1995) 
[E: Phys. Rev. D \textbf{55}, 5853 (1997)]. 

\bibitem{Bodwin07} 
G.T.~Bodwin, E.~Braaten, D.~Kang, J.~Lee, 
Phys. Rev. D \textbf{76}, 054001 (2007).  

\bibitem{cleo} 
G.~Viehhausser {\it et~al.}, 
{Nucl. Inst. Meth. A} \textbf{462}, 146 (2001).

\bibitem{chamb}
D.~Peterson {\it et~al.}, 
{Nucl. Inst. Meth. A} \textbf{478}, 142 (2002).

\bibitem{pid}
M.~Artuso {\it et~al.}, 
{Nucl. Inst. Meth. A} \textbf{502}, 91 (2003).

\bibitem{cryst}
Y.~Kubota {\it et~al.}, 
{Nucl. Inst. Meth. A} \textbf{320}, 66 (1992).

\bibitem{CLEOBerg}
S.~Kopp {\it et~al.} (CLEO Collaboration), 
{Phys. Rev. D} \textbf{63}, 092001 (2001).

\bibitem{mura} 
M.~Artuso {\it et~al.} (CLEO Collaboration), 
{Phys. Rev. Lett.} \textbf{94}, {032001} ({2005}).

\bibitem{cbline} 
T. Skwarnicki, Ph.D. Thesis, Cracow Institute of Nuclear Physics,  
DESY-F31-86-02, 1986; 
J.~Gaiser, Ph.D. Thesis, Stanford University, SLAC-255, 1982; 
R.~Lee, Ph.D. Thesis, Stanford University, SLAC-282, 1985.  

\bibitem{mw} 
This number is based on the work described in 
M.E.~Watkins, Ph.D. Thesis, Carnegie Mellon University, 2007 (unpublished), 
used here with a conservative error; 
it is also consistent with the predictions 
of D. Kang, T. Kim, J. Lee, C. Yu, Phys. Rev. {\bf D 76}, 114018 (2007).  

\bibitem{jetset} 
T.~Sj\"ostrand {\it et~al.}, 
Computer Physics Commun. \textbf{135}, 238 (2001).

\bibitem{BelleFrag} 
R.~Seuster {\it et~al.} (Belle Collaboration), 
Phys. Rev. D \textbf{73}, 032002 (2006). 

\bibitem{BelleContOnia} 
K.~Abe {\it et~al.} (Belle Collaboration), 
Phys. Rev. Lett. \textbf{88}, 052001 (2002). 

\bibitem{brambilla} 
N.~Brambilla, D.~Eiras, A.~Pineda, J.~Soto, A.~Vairo, 
Phys. Rev. Lett. {\bf 88}, 012003 (2001).  


\end{thebibliography}
\end{document}